\documentclass[a4paper,fleqn,usenatbib]{mnras}

\usepackage{newtxtext,newtxmath}

\usepackage[T1]{fontenc}
\usepackage{ae,aecompl}

\usepackage{graphicx}	
\usepackage{amsmath}	
\usepackage{amssymb}	

\title[The Intrinsic Far-UV Spectrum of B1422+231]
{The Intrinsic Far-UV Spectrum of the High-Redshift Quasar B1422$+$231}
\author[M. O'Dowd, N. F. Bate, R. L. Webster, K. Labrie, King, A. L., Yong, S-.Y.]
{M. O'Dowd$^{1,2,3}$\thanks{E-mail: matthew.odowd@lehman.cuny.edu}, 
N. F. Bate$^{4}$, 
R. L. Webster$^{5}$, 
K. Labrie$^6$, 
A. L. King$^{5}$, 
and S-.Y. Yong$^{5}$.\\
$^{1}$Department of Physics and Astronomy, Lehman College, City University of New York, \\
250 Bedford Park Boulevard West, Bronx, New York 10468-1589, USA\\
$^{2}$Department of Astrophysics, American Museum of Natural History, NY 10024-5192, USA\\
$^{3}$The Graduate Center of the City University of New York, 365 Fifth Avenue, New York, NY 10016, USA\\
$^{4}$Institute of Astronomy, University of Cambridge, Cambridge CB3 0HA, UK\\
$^{5}$School of Physics, University of Melbourne, Parkville, Victoria 3010, Australia\\
$^{6}$Gemini Observatory, Hilo, HI 96720, USA}

\begin{document}

\date{August 24, 2017}

\pubyear{Accepted}

\maketitle

\label{firstpage}

\begin{abstract}
We present new spectroscopy of the $z=3.62$ gravitationally lensed quasar B1422+117 from the Gemini North GMOS integral field spectrograph. We observe significant differential magnifications between the broad emission lines and the continuum, as well as across the velocity structure of the Lyman-$\alpha$ line. We take advantage of this differential microlensing to algebraically decompose the quasar spectrum into the absorbed broad emission line and absorbed continuum components. We use the latter to derive the intrinsic Ly$\alpha$ forest absorption spectrum. The proximity effect is clearly detected, with a proximity zone edge of 8.6--17.3~Mpc from the quasar, implying (perhaps intermittent) activity over at least 28~Myrs. The Ly$\alpha$ line profile exhibits a blue excess that is inconsistent with a symmetric fit to the unabsorbed red side. This has important implications for the use of this fitting technique in estimating the absorbed blue Ly$\alpha$ wings of Gunn-Peterson trough quasars. 
\end{abstract}


\section{Introduction}

The far ultraviolet (FUV) emission of luminous quasars is of great importance to our understanding of the evolution of the high redshift universe. The first quasars turned on in an intergalactic medium (IGM) of neutral hydrogen and may have competed with the first stars to reionize the universe \citep{Giallongo+15, MadauHaardt15, Finlator+16}. Those same quasars are now our most powerful probe of the epoch of reionization; Gunn-Peterson (GP) troughs in the spectra of $z\gtrsim 6$ quasars provide constraints on the residual neutral fraction at tail end of this process \citep{B01, W03, F06a, F06b, MesingerHaiman07, Mortlock+11,  Schroeder13}, and the size of quasars' ionization bubble, or proximity region, probes the IGM and provides a measure of quasars' energetic contribution to its evolution \citep{DonahueShull87, Bajtlik+88, WyitheLoeb04, F06b, MesingerHaiman07, BoltonHaehnelt07, WyitheBolton11, Schroeder13, Greig+17b}. The evolving state of the IGM at later times is then traced by the Lyman-$\alpha$ (Ly$\alpha$) forest in the  rest-frame FUV spectra of quasars at $z>1$ (See Weinberg et al. 2003 for a review). Although a wealth of information can be extracted from the absorption of quasars' intrinsic FUV spectrum, this same absorption means that this wavelength range is very poorly characterized. An improved understanding of the intrinsic FUV emission of luminous quasars is sorely needed.

Efforts to characterize the FUV spectrum of quasars are successful to a point; the use of spectroscopic samples to produce composite spectra \citep{Telfer+02, Shull+12, Stevans+14, BosmanBecker2015, Harris+16} or principle component analysis \citep{Suzuki+05, Suzuki06, LeeSpergel11} have given us insights into the true FUV continuum in the absorbed Lyman-$\alpha$ forest region. The Ly$\alpha$ broad emission line (BEL) itself is a greater challenge due to the wide range observed in BEL profile shapes, however progress has been made in understanding the systematics in fitting based on the unabsorbed Ly$\alpha$ red wing \citep{KramerHaiman09} and in reconstructing the line profile based on spectral correlations in other lines \citep{Greig+16}. However due to the ubiquitous absorption in the blue wing of the Ly$\alpha$ line at high redshifts, there remains a fundamental uncertainty in the range of possible Ly$\alpha$ emission line profile shapes in high-z quasars.

Gravitational lensing offers the potential to extract the elusive intrinsic FUV spectrum of high-z quasars. When a strongly-lensed quasar exhibits differential magnifications due to microlensing, it becomes possible to algebraically decompose the spectral contributions of different physical regions within the quasar \citep{Angonin90, Sluse07, O15}. This method has been used to extract intrinsic features within quasar spectra corresponding to particular physical processes and/or size-scales. It has also been used to extract the intrinsic broad absorption profile in the BAL quasar H1413+117 \citep{Angonin90, H10, O15, Sluse15}. In this paper we apply the technique to Gemini GMOS integral field spectroscopy of the $z=3.62$ quasar B1422$+$231. We derive the intrinsic Lyman forest absorption profile along the line of sight and constrain the intrinsic Ly$\alpha$ line shape and FUV spectrum of this quasar. 

In Section \ref{obs} we describe the observations and extraction of lensed image spectra. In Section \ref{sigs} we analyse the microlensing signatures in this new epoch of data.
In Section \ref{sd} we decompose the spectra in terms of the broad emission line region and continuum, emphasizing the physical interpretation of the decomposed regions, and we derive the intrinsic Lyman forest absorption profile. In Section \ref{intrinsic} we discuss the derived absorption spectrum and Ly$\alpha$ line shape. In Section \ref{conc} we summarize our conclusions.

\section{Observations and Spectral Extraction}
\label{obs}

B1422$+$231 was observed on the 27th of June 2006 with the GMOS
Integral Field Unit (IFU)
\citep{AllingtonSmith, Hook} on the Gemini North telescope. Five 15 minute exposures were taken in one-slit
mode using the B600 grating with a spectral range of 3975\AA\ to 6575\AA\ and a resolution of R$=$1688.
At B1422$+$231's redshift of 3.62, this covers the O{\sc vi}$+$Ly$\beta$ through C{\sc iii]}
broad emission lines.
Seeing conditions were excellent, with a PSF FWHM of 0\farcs6.

\subsection{Data Reduction and Spectral Exraction}

\begin{figure}
\includegraphics[width=75mm]{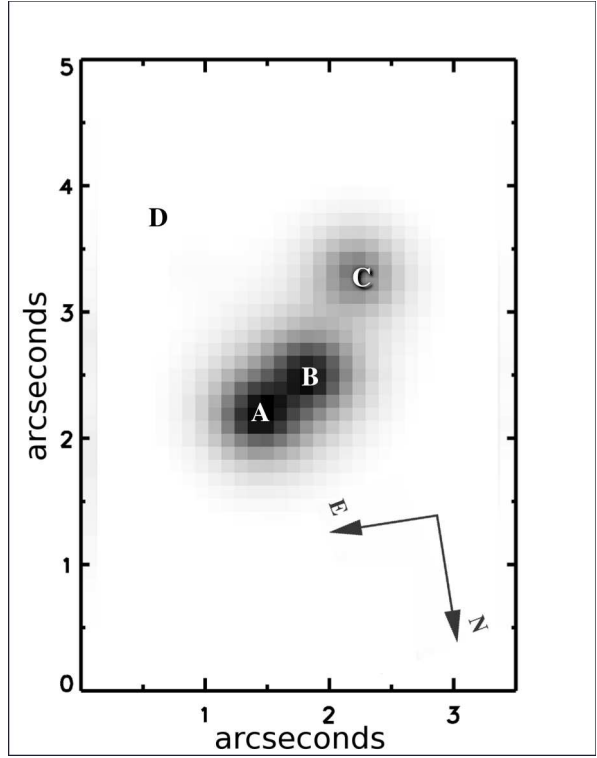}
  \caption{2-D image created from collapsing ADR-corrected IFU data
    cube. Letters indicate the standard designation of quasar images in B1422$+$231.}
  \label{2dimage}
\end{figure}

The data was reduced using the Gemini \textsc{IRAF} package v1.10,
under \textsc{IRAF} v2.14.1. See \citet{O15} for details of the reduction of this dataset, including the correction for scattered light. 

Figure \ref{2dimage} shows the processed data cube collapsed into a 2-D image.
The excellent seeing allows us to extract the spectra of lensed images C and
D with high accuracy, however images A and B, with their separation of
0$\farcs$5 require careful deblending. \citet{O11} describes the method used for the extraction of spectra for each lensed image, including deblending of close image pairs.

Images A, B, and C are distinct. Image D is faint but
confidently detected, however it is not visible in this grey-scale stretch. Images A and B are somewhat blended, but the
cores of their PSFs are resolved.

\begin{figure*}
\includegraphics[width=100mm]{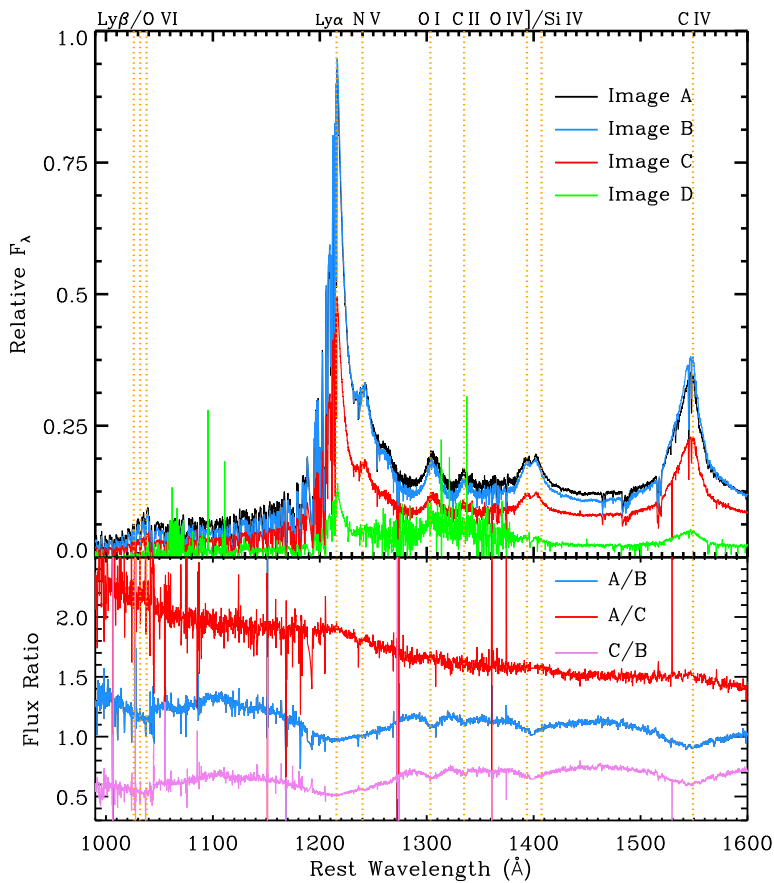}
  \caption{{\it Upper panel}: Extracted spectra for 4 lensed images in B1422$+$231. 
{\it Lower panel}: Ratios of spectra for lensed images (blue: A/B, red: A/C, violet: C/B). Residual features at locations of all broad emission lines for A/B and C/B reveal differential microlensing between the BELR and continuum. The relative absence of such features in A/C identifies B as the microlensed image.}
\label{spec}
\end{figure*}

Figure~\ref{spec} (upper panel) shows the extracted spectra of the four quasar images. 
We see distinct broad emission lines for O{\sc VI}$+$Ly$\beta$, Ly$\alpha$, N{\sc V},
O{\sc I}, C{\sc II}, Si{\sc IV}/O{\sc IV]}, and C{\sc IV}. 

\section{Microlensing Signatures}
\label{sigs}

Gravitational lensing is achromatic, and so in the absence of microlensing, differential extinction, or time-dependent spectral variation, the spectra of lensed images of a multiply-imaged quasar are identical modulo a scaling factor. Microlensing is typified by a difference in the magnification of one or more lensed images compared to the time-averaged magnification for that lensed image (which we will refer to as the macrolens magnification, or just macro-magnification). 
{\it Differential microlensing} is characterized by magnifications that vary between different emission regions, resulting in a change in the shape of the spectrum of the affected lensed image. Differential microlensing is sensitive to emission region size and projected location. It is frequently observed between BELs and the continuum, and often shows wavelength dependence within these features.

Figure~\ref{spec} (lower panel) shows the wavelength-dependent flux ratios of lensed image A with respect to B and C, and of C with respect to B. In the absence of microlensing we expect these to be flat. The A/C ratio is indeed relatively flat, although a constant slope is apparent that may result from differential microlensing across the continuum, however this is unlikely as we would then expect features in the ratio at the locations of the BELs, yet these are minimal. A more likely explanation is differential extinction, which has a smooth wavelength dependence. In A/B and C/B there are strong features at the locations of all BELs, indicating differential microlensing between the broad emission line region and the continuum. The relative absence of such features in the A/C ratio implies that lensed image B is experiencing this differential microlensing. 

Figure~\ref{continuum} shows the spectra of lensed images A and B in which both are scaled to have matching flux in the Ly$\alpha$ line core (rest frame 1216$\pm$20\AA). The continuum in B appears to be experiencing a lower magnification relative to the BELs. The BELs themselves also exhibit differential magnifications in lensed image B, with the wings of the lines at lower magnification relative to the cores. We discuss this in detail in \ref{fr}.

Differential extinction between lensed images will not affect the continuum-to-BEL ratios as our measurement of the continuum strength is made at the relevant BEL wavelength range (based on power-law fits; Sect~\ref{fr}), nor will it significantly affect broad line profiles. Differential extinction only affects the continuum slope and ratios between BELs, which are not a focus of his work.

In Section \ref{sd} we use the difference in continuum-to-BEL ratios between lensed images to decompose the spectrum into intrinsic components. It's important to note that for the purpose of the decomposition of spectral components, it does not matter why this ratio varies between images. The decomposition is equally valid whether these differences are due to microlensing or intrinsic variability, as long as wavelength-dependent variations within the BEL and continuum components can be accounted for. 

\subsection{Flux Ratios}
\label{fr}

\begin{figure}
  \includegraphics*[width=90mm]{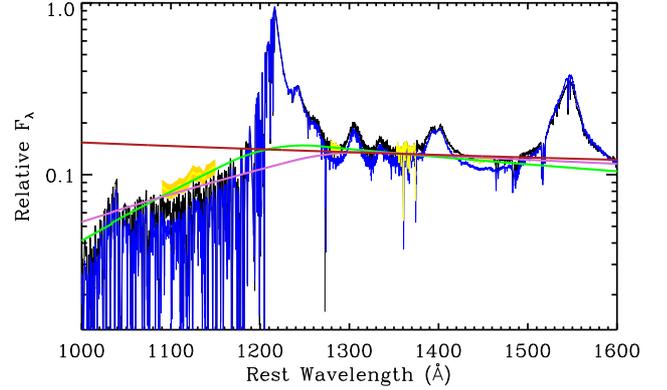}
  \caption{Spectrum of lensed image A (black) and B (blue) scaled to match Ly$\alpha$ flux at rest-frame 1186--1236\AA. Yellow regions redward of Ly$\alpha$ indicate regions of the spectrum of lensed image A (black) used for power-law continuum fits. The yellow region at rest-frame 1090-1150\AA\ shows the corrected ``de-absorbed'' Lyman-$\alpha$ forest used for the broken power-law continuum fit, along with the range of plausible intrinsic fluxes for the corrected continuum. Power-law fits for the lensed image A spectrum are shown; the dark red line is the best-fit unbroken power-law (fit redward of Ly$\alpha$; the green line is the best-fit broken power-law; the pink line shows a broken power-law fit with largest plausible break wavelength (1280\AA), chosen to allow the least change in spectral index. The full range of acceptible continuum fits are explored in the decomposition of Sect.~\ref{sd}.}
  \label{continuum}
\end{figure}

To inspect differential microlensing properties of B1422$+$231 we first obtain an estimate of the unabsorbed continuum spectrum for each lensed image. In the region redward of Ly$\alpha$, this can be done by fitting a power-law to the regions that are relatively uncontaminated by BELs. Fitting $F_\nu\propto \nu^{\alpha}$ in the windows 1280--1290\AA, 1350--1375\AA, and 1440--1460\AA\ yields $\alpha = -0.8\pm0.3$. This is similar to the mean 1200--2000\AA\ power-law index of $\alpha = -0.83$ for $z<1.5$ AGNs measured by \citet{Stevans+14}. The best fit single power-law continuum is shown in figure~\ref{continuum}. 

It is apparent that the best power-law fit redward of Ly$\alpha$ substantially overshoots the peaks of the Ly$\alpha$ forest. This may be due to from line blanketing, which at $z>3$ can be significant (Faucher-Gigu\'ere et al. 2008, hereafter FG08)  and/or a turnover in the extreme UV (EUV) spectrum. Although quasar spectra often exhibit EUV turnovers, these are typically gradual and blueward of $\sim$1000--1100\AA\ \citep{Shull+12, Stevans+14}. 

We can estimate the contribution of these effects by correcting the observed EUV spectrum for the average IGM absorption.  To estimate the latter, we use the $2<z<4$ average IGM opacity measurement of FG08. We calculate the expected continuum level in the rest-frame 1090-1150\AA\ range, which is reasonably uncontaminated by BELs. The uncertainty in our unabsorbed continuum estimate comes from the uncertainty in the FG08 fits and the standard error of the mean in the smoothed Ly$\alpha$ forest. Figure~\ref{continuum} shows the derived absorption-corrected spectrum in the 1090-1150\AA\ range. The corrected EUV continuum remains significantly below the extrapolated single-index continuum. This may indicate that B1422$+$231 exhibits an unusually long-wavelength spectra index break, or that our estimate of line blanketing is low. To study the full range of possible continuum shapes we perform our analysis assuming both an unbroken power-law and a broken power-law fit to the opacity-corrected EUV continuum. 

For the latter we stitch the above $\alpha = -0.8\pm0.3$ power law redward of Ly$\alpha$ to a power-law fit at 1090-1150\AA, which yields $\alpha = -2.0\pm0.5$. This is steep compared to the mean 500---1000\AA\ power-law index of $\alpha = -1.41$ measured by \citet{Stevans+14}, however it is within the range of EUV slopes found in that study. Additional free parameters in deriving the broken power-law continuum were the width of a Gaussian smoothing function over the break region and the break wavelength. These latter have very little effect on the resulting analysis. Figure~\ref{continuum} shows the best-fit broken power-law. We use both the unbroken and broken power-law continua spectra to explore the range of spectral decompositions in Section~\ref{sd}.

Table~\ref{fluxratios} gives the flux ratios for the Ly$\alpha$, O{\sc VI}$+$Ly$\beta$, O{\sc I}, O{\sc II}, O{\sc IV]}$+$Si{\sc IV}, and C{\sc IV} 
emission lines and for the continuum. For all line ratios but Ly$\alpha$, integrated, continuum-subtracted line fluxes are calculated over windows centered on rest-frame emission line wavelengths with widths given in Table~\ref{fluxratios} . For Ly$\alpha$ we calculate the flux ratio in the 10\AA\ window redward of rest-frame 1216\AA\ and blueward of significant N{\sc V} emission.
\begin{table}
\caption{Measured flux ratios for broad emission lines and continuum. 8~GHz flux ratios from \citet{Patnaik+99}  and mid-IR flux ratios from \citet{Chiba05} are included.}
\label{fluxratios}
\begin{tabular}{lcccc}
\hline
\hline
      & A/B & A/C & C/B &$\Delta\lambda$(\AA)  \\
\hline
Ly$\alpha$              &  0.93 $\pm$ 0.02 & 1.92 $\pm$ 0.05 &  0.48 $\pm$ 0.02  &10\\  
O{\sc VI} $+$Ly$\beta$               &  1.15 $\pm$ 0.40 &  2.25 $\pm$ 0.45 & 0.61 $\pm$  0.20  &15\\ 
O{\sc I}                    & 0.92 $\pm$ 0.10 & 1.79 $\pm$ 0.35 &  0.51 $\pm$ 0.07   &15\\  
C{\sc II}                   & 0.95 $\pm$ 0.20 & 1.79 $\pm$ 0.79 &  0.53 $\pm$ 0.18   &10\\  
O{\sc IV]}$+$Si{\sc IV}   &0.93 $\pm$ 0.06 & 1.64 $\pm$ 0.14 & 0.56 $\pm$ 0.04    &30\\  
C{\sc IV}                  & 0.84 $\pm$ 0.04 & 1.53 $\pm$ 0.07 &  0.55 $\pm$ 0.02    &40\\  
\hline
Continuum & 1.16 $\pm$ 0.01 & 1.64 $\pm$ 0.03 & 1.42 $\pm$ 0.04 &\\
\hline
MIR                         & 0.94  $\pm$ 0.05& & 0.57  $\pm$ 0.06&\\
8~GHz                     & 0.93 $\pm$ 0.02 & 1.89 $\pm$ 0.02  &0.49 $\pm$ 0.01&\\
\hline
\hline
\end{tabular}
\end{table}

\begin{figure}
  \includegraphics*[width=75mm]{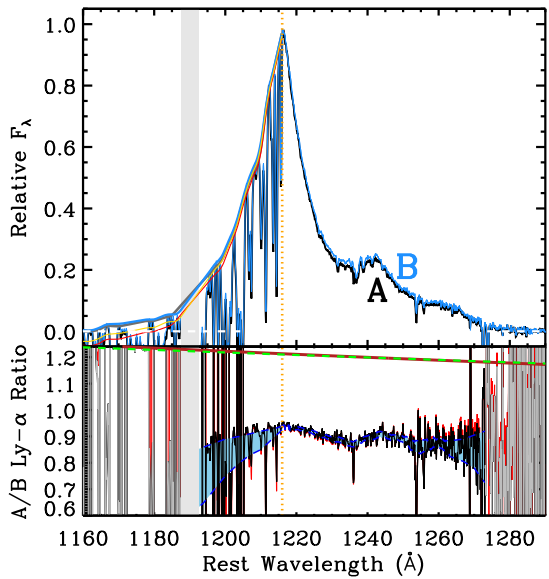}
  \caption{{\it Upper panel}: Ly$\alpha$ line for lensed images A and B with best-fit power-law continua subtracted. Best broken power-law continua subtractions are shown in black (image A) and blue (image B). For clarity, lines are draw connecting the peaks of the Ly$\alpha$ forest, and these also show the unbroken power-law subtractions in red (image A) and yellow (image B). {\it Lower panel}: Continuum-subtracted Ly$\alpha$ A/B line ratio (broken power-law subtraction in black, unbroken in red). The blue dashed lines show the range of possible A/B ratios based on propogating both noise and the uncertainties in the continuum fits (including the range from broken to unbroken power-laws). 
The dark red (dashed green) line shows the A/B continuum ratio based on the broken (unbroken) power-law continuum fits. The grey region indicates the location of the chip gap; data is absent in this region.}
\label{lineandratio}
\end{figure}

\begin{figure}
  \includegraphics*[width=75mm]{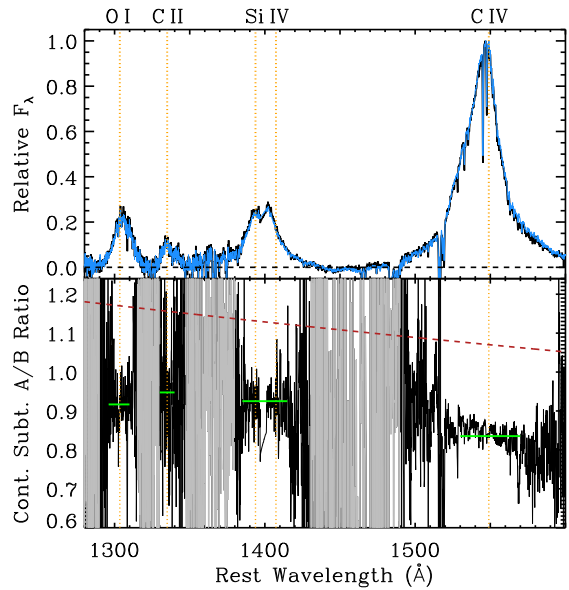}
  \caption{{\it Upper panel}: O{\sc I}, C{\sc II}, Si{\sc IV}, and C{\sc IV} BELs for lensed images A (black) and B (blue) with best-fit continua subtracted. {\it Lower panel}: Continuum-subtracted A/B line ratios. Green lines show the mean flux ratios for each BEL along with the range over which those ratios were taken.
The dark red line shows the A/B continuum ratio based on the continuum fits.
}
\label{otherlineandratio}
\end{figure}

Figure~\ref{lineandratio} shows the continuum-subtracted Ly$\alpha$ line for lensed images A and B, along with the A/B line ratio taken after continuum subtraction. This ratio reveals clear differential microlensing across the velocity structure of the Ly$\alpha$ line. The line center has a microlensed magnfication ratio that is closer to the continuum ratio (also shown) compared to the red and blue wings. The NV line shows a similar behavior. Figure~\ref{otherlineandratio} shows the continuum-subtracted spectrum spanning the remaining BEL species O{\sc I}, C{\sc II}, Si{\sc IV}, and C{\sc IV}, including the continuum-subtracted A/B ratio. The range of uncertainties n this ratio are also shown in Fig.~\ref{lineandratio}, based on propagating the uncertainties due to shot noise and the continuum fit. 

\begin{figure}
  \includegraphics*[width=60mm, angle=-90, origin=c]{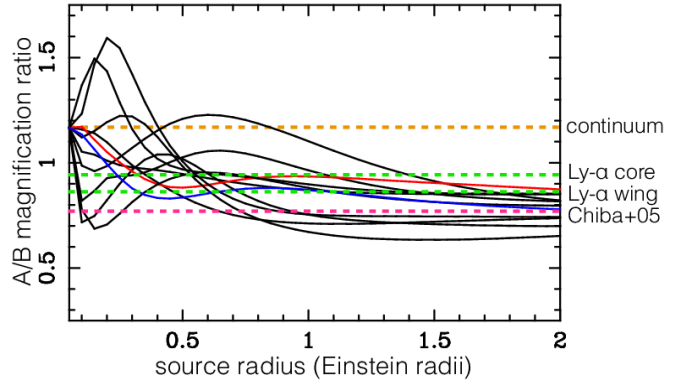}
  \caption{Simulated A/B magnification ratios as a function of Gaussian source size. 11 random location pairs are chosen on simulated magnification maps for lensed images A and B, with locations constrained to match the magnification ratio observed in the continuum. {\it Dark red dashed line} shows the average continuum magnification ratio; {\it orange dashed lines} shows the average Ly$\alpha$ core and wing ratios; {\it black dashed line} shows the 11.7~$\mu$m flux ratio from \citet{Chiba05}. Several tracks can be seen to vary non-monotonically with increasing source size. The red and blue lines are examples of tracks in which the high-velocity Ly$\alpha$ wing is consistent with having a smaller source size than the lower-velocity line core, even though that wing has a flux ratio closer to the expected large-scale flux ratio.
}
\label{convergence}
\end{figure}

A naive interpretation of these relative magnification ratios suggests that the low-projected-velocity core of the Ly$\alpha$ line may be closer in scale to the continuum emission than the higher velocity wings. Such an increase in BELR velocity with increasing size crudely suggests an accelerating outflow, and that we are seeing little orbital motion towards the base of this flow. 
However we must be cautious before drawing conclusions about the kinematics of the Ly$\alpha$ line. Magnification ratio does not always tend towards the macromodel ratio monotonically with increasing source size in the event of size-dependent differential microlensing. This is demonstrated in figure~\ref{convergence}, which shows a random selection of A/B magnification ratios simulated for B1422$+$231 for source sizes between 0.1--2 Einstein Radii. In this simulation, Gaussian surface brightness profiles over this size range are convolved with simulated microlensing magnification maps produced as part of the GERLUMPH project \citep{gerlumph}. We use the B1422$+$231 lens macro-model of \citet{Mediavilla09} with an 80\% smooth matter percentage, consistent with the values estimated in other lenses where the line of sight does not pass through the galactic core \citep{Keeton06, Chartas09, Pooley09, Dai09, Bate11}. Tracks show how the expected magnification ratios change with source size for a given set of locations for images A \& B on the simulated microlensing magnification maps. For a number of tracks our observed magnification ratios are consistent with the high velocity wings of Ly$\alpha$ arising from a smaller size scale than the low velocity core.

\section{Spectral Decomposition}
\label{sd}

Gravitational microlensing grants the ability to
decompose quasar spectra, separating components that are subject
to different levels of microlensing. 
The basic approach is to express the spectra of two lensed images as
linear combinations of differently-magnified components. These 
equations can then be solved for an unknown spectral component.
The technique has been used to decompose microlensed from non-microlensed 
components (e.g. the $F\mu$ method --- \citealt{Sluse07, H10, Sluse11}), as well as extract 
what can be interpreted as intrinsic broad line, continuum components, and absorption features
\citep{Angonin90, H10, O15}. 

In \citet{O15} we describe our approach to spectral decomposition, and we refer the reader there for more detail.
In short, we express the observed spectrum as a linear combination of an intrinsic continuum spectrum, $F_C(\lambda)$, and broad emission line spectrum, $F_L(\lambda)$. Each is subject to its own magnification factor, $m_C(\lambda)$ and $m_L(\lambda)$,
which describe both macro- and microlensing magnifications, and which may or may not have wavelength dependence.
These factors may also encompass any other differential factors between these spectral components, for example due to intrinsic 
variability. We also describe a separate absorption factor for continuum and BEL, $A_C(\lambda)$ and $A_L(\lambda)$. 
In \citet{O15}  this absorption component described the intrinsic broad absorption features in H1413$+$117. In the case of B1422$+$231 they primarily describe the Lyman forest absorption.
The spectra for images A and B then become:

\begin{equation}
{\bf F}_A = {\bf m}_{C,A} {\bf A}_C {\bf F}_C + {\bf m}_{L,A} {\bf A}_L {\bf F}_L\\
\end{equation}

\begin{equation}
 {\bf F}_B = {\bf m}_{C,B} {\bf A}_C {\bf F}_C + {\bf m}_{L,B} {\bf A}_L {\bf F}_L
\end{equation}

Bold symbols represent potential wavelength dependence. The only known quantities in these equations are the observed spectra, \textbf{\textit{F}}$_A$ and \textbf{\textit{F}}$_B$, however it is possible to estimate the magnified continuum spectrum for each lensed image (${\bf m}_C {\bf F}_C$;  see Sect.~\ref{fr}) and the continuum and BEL flux ratios, which we define as ${\boldsymbol\eta}_C = {\bf m}_{C,A}/{\bf m}_{C,B}$ and ${\boldsymbol\eta}_L$ = \textbf{\textit{m}}$_{L,A}/$\textbf{\textit{m}}$_{L,B}$. 

We can then solve for the {\it absorbed}, magnified continuum $\mathcal C$ and BEL $\mathcal L$ spectrum for either lensed image:

\begin{equation}
\label{belr}
\boldsymbol{\mathcal L}_A = {\bf m}_{L,A} {\bf A}_L {\bf F}_L = \frac{{\boldsymbol\eta}_L}{{\boldsymbol\eta}_L - {\boldsymbol\eta}_C} ({\bf F}_A - {\boldsymbol\eta}_C {\bf F}_B)
\end{equation}

\begin{equation}
\label{cont}
\boldsymbol{\mathcal C}_A = {\bf m}_{C,A} {\bf A}_C {\bf F}_C = \frac{{\boldsymbol\eta}_C}{{\boldsymbol\eta}_C - {\boldsymbol\eta}_L} ({\bf F}_A - {\boldsymbol\eta}_L {\bf F}_B)
\end{equation}

\begin{figure}
  \includegraphics*[width=85mm]{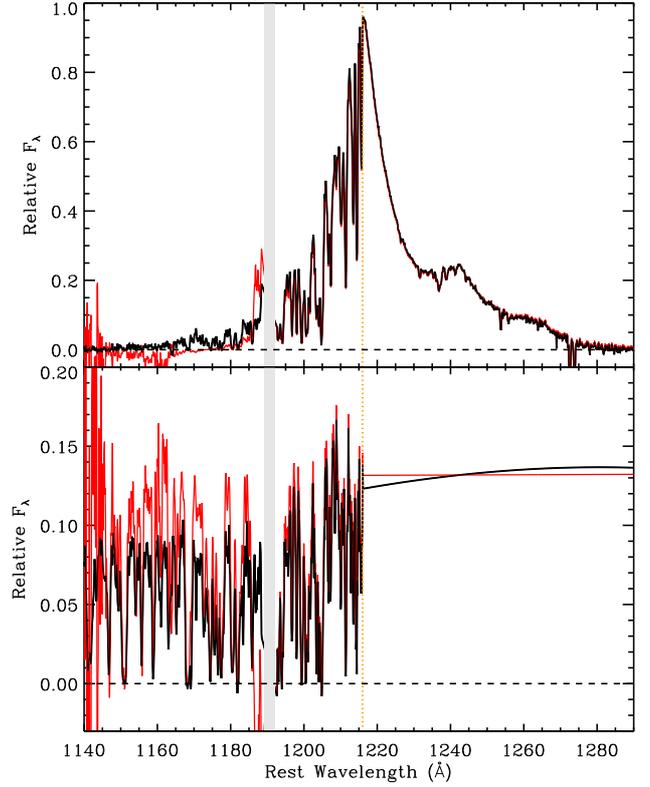}
  \caption{Spectral components derived from decomposition of B1422$+$231 in the wavelength range of the Ly$\alpha$ broad emission line. The {\it upper panel} shows the absorbed, magnified Ly$\alpha$ line, while the {\it lower panel} shows the absorbed, magnified continuum component in this range. Black curves show decompositions assuming the best-fit broken power-law continuum, while red curves assume an unbroken power-law continuum fit redwards of Ly$\alpha$. 
Rest wavelengths are determined assuming $z=3.62$. The orange dashed line shows the rest wavelength of Ly$\alpha$. 
 The grey bar represents the location of the chip gap; data is absent in this region.}
\label{decomp}
\end{figure}

We take ${\boldsymbol\eta}_C$ to be the wavelength-dependent continuum ratio from the fits described in Section~\ref{fr}. As seen in Sect.\ref{fr}, ${\boldsymbol\eta}_L$ is also wavelength dependent for Ly$\alpha$, and so we use the continuum subtracted line ratio for this component. 
Figure~\ref{decomp} shows the $\boldsymbol{\mathcal L}_A$ and $\boldsymbol{\mathcal C}_A$ components derived in the region of the Ly$\alpha$ line from equations \ref{belr} and \ref{cont} assuming the best-fit continua (both broken and unbroken power-law fits). We note that the decompositions assuming an unbroken power-law fit for the continuum produce significantly negative fluxes in the line component below rest-frame $\sim$1160\AA, and below 1145\AA\ results in extreme fluctuations in both decompositions. We take this as an indication that there is at least some spectal index turnover in the continuum at or above $\sim$1160\AA.

\section{Estimating the Intrinsic Absorption and Line Profiles}
\label{intrinsic}

Following reionization, quasars continued to contribute a significant part of the energy budget of the IGM. All quasars at z$>$1 are surrounded by a region of diminished opacity due to their ionization of residual neutral hydrogen clouds in the IGM. The size-scale of this proximity zone is typically taken to be the distance from the quasar at which the quasar's ionizing flux equals the metagalactic background \citep{ShapiroGiroux1987, CenHaiman2000}. However, in practice, the duration of a quasar's activity also plays a part, and defines the far edge of the ionization front. Measuring the exact size and shape of proximity region absorption is very challenging because it overlaps the quasar's own Ly$\alpha$ broad emission line, and the profile of the Ly$\alpha$ blue wing in these luminous high-z quasars has never been observed due to this same absorption.
  
\subsection{The Intrinsic Lyman-$\alpha$ Forest}
\label{lyforest}

\begin{figure}
  \includegraphics*[width=85mm]{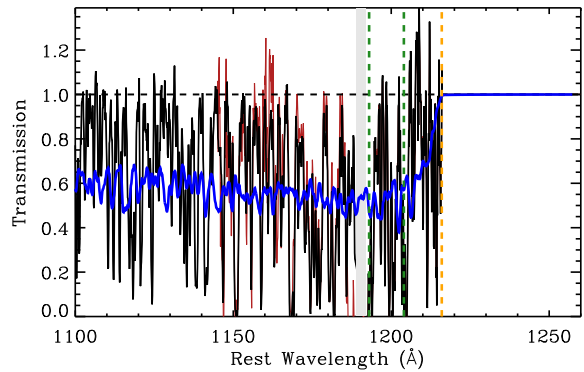}
  \caption{The absorption profile experienced by continuum light traveling the path of lensed image A. The black curve is the absorption profile assuming the best-fit broken power-law continuum, while the red curve assumes the best-fit to unbroken power-law continuum fit redward of Ly$\alpha$. These overlap almost perfectly in the proximity zone. The unbroken continuum profile is cut off below 1145\AA\ as this model predicts significantly negative line flux and the decomposition is poorly contrained.
The blue line shows the mean transmission spectrum for 50 sightlines towards a $z=3.62$, $M_V=-26.8$ quasar added to the C15 hydrodynamic simulation of \citet{Becker+11} and scaled to the opacity measurements of FG08. Vertical dashed lines show the rest wavelength of Ly$\alpha$ (orange) and the approximate lower and upper limits on the wavelength extent of the proximity region (black) used for estimating the physical proximity region size in Sect.~\ref{lyforest}. The grey bar represents the location of the chip gap; data is absent in this region.}
\label{intabs}
\end{figure}

The derived continuum component reveals the intrinsic shape of the Ly$\alpha$ absorption spectrum  (Fig.~\ref{decomp}, lower). By dividing $\boldsymbol{\mathcal C}_A$ by the continuum spectrum fit for lensed image A, $m_{C,A} F_C$ we recover $A_{C,A}$, the transmission function experienced by the continuum light traveling the line-of-sight path of lensed image A ($A_{C,A}$), as shown in Figure~\ref{intabs}. 

Certain features are clearly independent of these uncertainties:

1) The onset of Lyman forest absorption is coincident with the Ly$\alpha$ line center, indicating no significant infalling HI contributing to the absorption.

2) A region of enhanced transmission in the vicinity of the quasar is clear, and presumably results from the line-of-sight (LOS) proximity effect within the quasar's Str\"omgren sphere. Transmission drops off blueward of the resonant Ly$\alpha$ wavelength, reaching the IGM average in the observed-frame wavelength range $\sim$5560---5500\AA. This corresponds resonant Ly$\alpha$ absorption at z$\sim$3.57---3.52, or a proper distance of 8.6--17.3~Mpc assuming $H_0 = 70~\text{km s}^{-1}\text{Mpc}^{-1}$, $\Omega_M = 0.3$ and $\Omega_\Lambda = 0.7$. Note that this proper distance corresponds to the observable edge of the proximity zone. This is distinct from the {\it characteristic size} of this zone, typically defined as the distance at which the ionizing flux of the quasar is equal to metagalactic ionizing flux. Instead, we estimate the distance at which the opacity approaches the average IGM value at that redshift, corresponding to the point at which the ionization rate due to the quasar is small compared to that due to the metagalactic background flux. From light travel time arguments alone, this radius may be taken as a lower limit on the timescale of the quasar's activity. As such, B1422$+$231 appears to have been luminous, at least intermittently, for more than 28~Myrs. 

This activity timescale is comparible to the $>$34~Myr timescale for the $z=3.286$ quasar Q0302$-$003 based on its He{\sc II} proximity zone \citep{SyphersShull2014}. These authors also determine that an integrated activity period of as little as 0.17 Myrs is sufficient to ionize this zone, implying intermittent activity and a small duty cycle. 

To illustrate how an intermittently luminous quasar can produce a proximity effect comparible to that observed for B1422$+$231, we have generated a set of simulated absorption profiles using the C15 hydrodynamic simulation of \citet{Becker+11}. This was scaled to reproduce the mean opacity of FG10, and includes the ionizing effect of continuously shining quasar at $z=3.62$ with the average quasar spectrum of \citet{Madau+99} (James Bolton, private communication). B1422$+$231's unlensed absolute magnitude is calculated assuming $\alpha=-0.8$, $m_V= 16.4$---$16.7$ for lensed image A (\citet{Remy+93, YeeEllington94}; CASTLES survey\footnote{https://www.cfa.harvard.edu/castles}), and a lensing magnification of $\mu_A\sim 5$---10; \citep{Kormann+94, Mediavilla09, Schechter+14}. B1422$+$231 is extremely luminous, with $M_V \sim 28.6$ to 29.6. To approximate the effect of intermittent activity with a duty cycle of 0.1 to 0.2, we simulate the absorption profile using a continuously-emitting quasar with $M_V=-26.8$, so 5 to 10 times fainter than B1422$+$231. Figure~\ref{intabs} shows the average of 50 simulated sight-lines towards the quasar. It can be seen that this power output results in a region of decreased opacity with a similar size to the proximity zone of B1422$+$231. Note that this simulation is for illustrative purposes only, and is not intented to provide constraints on any model parameters. Nonetheless, the duty cycle needn't be higher than 0.1 to produce the observed proximity effect.

3) Outside the proximity region the transmission spectrum rises relatively monotonically towards lower redshifts. This is as expected, and is reasonably consistent with the opacity-redshift relation of FG08. Note that this derived absorption profile uses a single opacity from FG08 (the average in the z$\sim$3.15---3.35 range) as an input parameter in our estimate of the intrinsic UV continuum. However this is effectively just a scaling factor; the monotonic increase in transmission is a result of the decomposition and a feature of the data. The absorption profile derived assuming a broken power-law continuum  is consistent with the simulated profile, while the profile derived from the unbroken power-law rises alittle too quickly.

\subsection{The Intrinsic Lyman-$\alpha$ Emission Line Profile}
\label{intline}

\begin{figure}
  \includegraphics*[width=95mm]{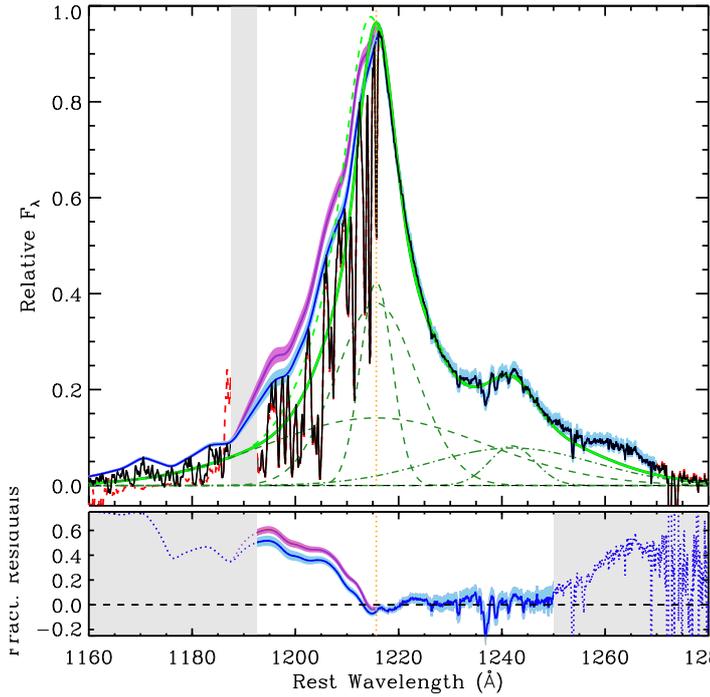}
  \caption{{\it Upper panel}: Derived Ly$\alpha$ broad emission line for lensed image A, subject to Ly$\alpha$ forest absorption and lensing magnification ({\it black line}) at rest-frame wavelength (orange dotted line). The envelope of the Ly$\alpha$ forest ({\it blue line}) indicates the minimum flux in the blue wing. The {\it red line} estimates the intrinsic blue-wing flux assuming some opacity at the transmission peaks based on an estimate of the minimum opacity outside the near zone (see Sect.~\ref{intline} for details). The {\it blue and red shaded areas} around these lines are the uncertainty intervals based on propagating continuum fit uncertainties through the decomposition process. The {\it solid green line} is the best composite Gaussian fit to 1216\AA\ $<\lambda < 1250$\AA\ range with fixed central wavelength (1215.67\AA); the {\it dashed green line} allows the Ly$\alpha$ central wavelength as a free parameter (line center: 1214.5\AA). The {\it grey shaded region} indicates the location of the CCD chip gap. {\it Lower panel}: The difference between the best-fit, fixed-wavelength composite Gaussian and derived Ly$\alpha$ line, with the blue wing taken as the Ly$\alpha$ forest envelope ({\it blue line}) and the absorption-corrected envelope ({\it red line}). {\it Blue and red shaded areas} are the uncertainty intervals. {\it Shaded grey regions} indicate where the fit is not of interest due to low flux (left) or the influence of unfitted emission line components (right).}
\label{gaussian}
\end{figure}

Figure~\ref{gaussian} shows the derived broad emission line component over the range of the Ly$\alpha$ line. It's important to remember that this is component is the combination $m_{L,A} A_L F_L$: the intrinsic line profile ($F_L$) as affected by Ly$\alpha$ forest absorption over the footprint of the broad line region ($A_L$) and the size-dependent magnification factor for Ly$\alpha$ at lensed image A ($m_{L,A}$). However, even with these convoluting influences, we can derive useful contraints on the intrinsic line profile.

The line connecting the peaks of the Ly$\alpha$ forest can be taken as a minimum for the intrinsic flux of the blue wing of the Ly$\alpha$ line; i.e. assuming no absorption in the regions corresponding to these peaks. In Section~\ref{fr} we saw that, based on the IGM opacity measurements of FG08, some absorption is likely even at the Ly$\alpha$ forest peaks for the region blueward of the Ly$\alpha$ line (Fig. \ref{continuum}). Average opacity, and also opacity at the forest peaks will diminish approaching the Ly$\alpha$ line center due to a decreasing neutral fraction inside the quasar's ionization front. We use the formalism of \citet{Calverley+11} to describe the relationship between opacity and distance from the quasar, assuming the range of proximity zone edge sizes (8.6---17.3~Mpc) measured in Section~\ref{lyforest}, and assuming that the measured edge size corresponds to the point at which opacity is within 10\% of the average IGM forest opacity. This proximity region opacity is scaled to converge with the IGM opacity in the regions corresponding to the Ly$\alpha$ forest peaks. In this way we can correct the upper envelope of the Ly$\alpha$ forest to produce an ``intrinsic'' Ly$\alpha$ line profile, also shown in Figure~\ref{gaussian}.

The shaded areas surrounding this ``intrinsic'' line profile and the Ly$\alpha$ forest peaks show the range of possible reconstructed line profiles based on propagating the uncertainties in the continuum fit (Sect.~\ref{fr}) through the decomposition. These include: uncertainties in spectral indices, normalization, continuum break wavelength, and the width of the Gaussian smoothing kernel across that break. Note that these fit uncertainties incorporate the uncertainty in the absorption correction for the FUV continuum. The uncertainty in the continuum normalization dominates over most of the blue wing of the line profile. The uncertainty in the absorption-corrected profile (red region) includes the range of proximity zone sizes measured in Section~\ref{lyforest}.

\subsubsection{Ly$\alpha$ Blue Excess and Implications for Reionization Studies}

The derived Ly$\alpha$ component enables us to place the first strong constraints on the intrinsic blue wing of the Ly$\alpha$ line profile in a luminous, high-z quasar. Understanding the Ly$\alpha$ profile, and in particular its potential asymmetries, is critical for reionization studies. Tracking the final stages of reionization depends heavily on measuring the depths of Gunn-Peterson (GP) troughs \citep{GunnPeterson65}, GP damping wings \citep{ME98}, and proximity region sizes for quasars at $z\gtrsim6$. For these measurements the blue wing of the Ly$\alpha$ line must be inferred, and its intrinsic shape is a key uncertainty in derived neutral fractions. A common approach is to fit an analytic profile to the unabsorbed red wing, for example Voigt profiles \citep{Schroeder13}, composite Gaussians \citep{F06b, MesingerHaiman07, Carilli10}, while a recent more sophisticated approach by \citet{Greig+16} tunes a composite Gaussian to variations in longer wavelength lines. All of these approaches assume symmetry in the Ly$\alpha$ line profile. Another other common approach is to construct composite spectra from lower redshift quasars with unabsorbed Ly$\alpha$ lines \citep{F06a, Mortlock+11}, however the largely $z\lesssim 1$ quasars necessarily used for these composites are for the most part very different in power to the extremely luminous quasars observed at $z\gtrsim 6$.

Ly$\alpha$ line profiles in low-to-medium redshift quasars do indeed exhibit asymmetries. \citet{KramerHaiman09} (KH09) analyzed Ly$\alpha$ profiles from archival HST spectroscopy of 78 quasars at $z\lesssim 1.5$, finding that Gaussian-fitted narrow (broad) line component have median blueshifts relative to the Ly$\alpha$ rest frame of $\sim$300 ($\sim$400~km/s)~km/s. KH09 analyze the effect of Ly$\alpha$ blue excess on reionization studies; they simulate $z>6$ GP troughs in their low-z sample and attempt to extract neutral fractions by fitting symmetric Ly$\alpha$ line profiles, with line centers based on the metal lines. They find that neutral fraction would be systematically underestimated if high-z quasars had similar Ly$\alpha$ profile asymmetries as are observed at low z. While this sample may not be representative of the typically much more luminous quasars used as reionization probes, the result advocates caution in how neutral fraction measurements are interpreted. This result also motivates us to determine whether luminous, high-z quasars have similarly asymmetric Ly$\alpha$ to their lower luminosity counterparts at low z.

At first glance, the derived Ly$\alpha$ line profile for H1413+117 is inded asymmetric around the Ly$\alpha$ rest wavelength; it appears to have some blue excess based on the flux at the Ly$\alpha$ forest peaks. 
We assess this excess by fitting a symmetric function to the red wing of the derived profile at $1216\AA <\lambda < 1250\AA$. We deliberately ``over-fit'' with three independent Gaussians at the Ly$\alpha$ line center and two Gaussians at N$\sc{V}$. Our aim is to assess the implication of symmetric reflection of the red wing in the case of an excellent fit,  and in the case of this Ly$\alpha$ profile we don't get an adequate fit with a double Gaussian. Note, however, that any fitted symmetric function will show a blue excess. At first we fix centers to systemic rest-frame wavelengths and fit normalizations as free parameters. The resulting best fit composite Gaussian is shown in Figure~\ref{gaussian}. The fit falls significantly below the peaks of the Lyman-$\alpha$ forest in the blue wing of the Ly$\alpha$ line. To check whether a blue-shifted central Ly$\alpha$ line wavelength can reproduce the apparent blue excess without a line asymmetry, we attempt the same fit with central wavelength also a free parameter. The fits are poorly constrained, but in general it is impossible to reproduce the blue excess in the high-velocity broad component without significantly overshooting flux in the narrow component core. This would imply an opacity that increases close to the quasar, yet we expect and only ever observe the reverse of this. A typical fit with a blue-shifted line center is also shown in Figure~\ref{gaussian}.

We conclude that a symmetric fit falls significantly below the true line flux in the blue wing. The implied blue excess may result from gas kinematics (outflow) or contributions from other emission lines such as Si{\sc III} (1206\AA), Si{\sc II} (1190, 1193\AA),  and C{\sc  III} (1175\AA). 
 
The lower panel of Figure~\ref{gaussian} shows the fractional residuals, comparing the best-fit composite Gaussian (fixed Ly$\alpha$ wavelength) to two cases: 1) the envelope of the Ly$\alpha$ forest line peaks (the minimum blue-wing flux), and 2) the absorption corrected blue wing. The mismatch rises to at least $\sim$33\% from line center to 10\AA\ from the core, and the true line flux is anywhere between this and 60\% higher than the fit across the blue wing. If a similar blue excess were present in a $z\gtrsim 6$ GP trough quasar then it would lead to a significant underestimate of the neutral fraction. This underestimate would be most significant for neutral fraction measures that use the GP trough itself. A measurement using the GP damping wing would be less affected as this probes much closer to the Ly$\alpha$ core; typically within rest-frame 10\AA\ \citep{MesingerHaiman07}, where the blue excess in B1422+231 is less than 20\%. 

Combined with the results of KH09, this clear Ly$\alpha$ blue excess in a powerful, high-z quasar suggests that neutral fractions may be systematically underestimated in many studies. As KH09 point out, in studies affected by this bias we may interpret any detection of a significant neutral fraction even more confidently.

\section{Conclusions}
\label{conc}
We have presented new spectroscopy of the gravitationally lensed quasar B1422+117 using Gemini North's GMOS IFU. Extracted spectra of the four lensed images reveal that differential microlensing strongly affects the relative magnification of the broad emission lines compared to the continuum in lensed image B, with the continuum in B appearing anomalously bright compared to all other lensed images. Differential microlensing is apparent across the velocity structure of the Ly$\alpha$ line, in which the line center of lensed image B is more strongly microlensed; it is closer to the magnification of the continuum, while the higher velocity wings appears to be closer to the expected macro-model magnification. This does not enable trivial constraints on the kinematics due to the non-monotonic dependence of microlensing magnification on source size, however with more detailed simulations such contraints may be possible.

We take advantage of the differential magnification between lensed images A and B to algebraically decompose the quasar spectrum into two components: the absorbed broad emission line spectrum and the absorbed continuum spectrum, and we use the latter to derive the intrinsic Lyman forest absorption profile. Analysis of these components leads to the following conclusions:

\noindent
{\bf $\bullet$}
 The proximity zone of the B1422+117 has an outer edge at a proper distance of between 8.6--17.3~Mpc from the quasar, indicating activity over at least 28~Myrs. Note that this does not necessarily mean a period of continuous activity of this length; a duty cycle of 0.1 is more than adequate to produce a similar proximity zone.

\noindent
{\bf $\bullet$}
The monotonic increase in IGM opacity with increasing redshift is consistent with that measured for larger quasar samples.

\noindent
{\bf $\bullet$}
A Gaussian fit to the red wing shows that the blue wing of the reconstructed Ly$\alpha$ line exhibits significant excess flux relative to that fit. This indicates that caution should be taken in the interpreting neutral fraction measurements based on GP trough quasar Ly$\alpha$ lines. If the blue excess observed in the  B1422$+$231's Ly$\alpha$ line is typical for luminous, high-z quasars then the common assumption of a symmetric line profile may have lead to significant underestimates of the neutral fraction.

\section*{Acknowledgements}
We would like to thank Michael Shull for his illuminating referee report which helped us to dramatically improve and clarify this work. We are also very grateful to James Bolton for generating tailored hydrodynamic simulations of the absorption profile. Finally, we thank the Gemini Observatory for their collaboration in acquiring and reducing these data.

\bsp

\label{lastpage}

\end{document}